# PHOTON-DETECTOR ENTANGLEMENT BY INTERFERENCE AT BEAM SPLITTER

Vladan Panković

Department of Physics, Faculty of Sciences, 21000 Novi Sad, Trg Dositeja Obradovića 4
Serbia, vladan.pankovic@df.uns.ac.rs

## Abstract

In early days of navigation over sea and ocean it has been considered that when a ship, governed by waves, arrives as some critical distance of the global character, then this ship must absolute collapse, i.e. must fall in the nothing. Only remarkable Magellan voyage definitely "experimentally" realized proof that there is no absolute collapse of the ship trajectory on the ocean, that different collapses of the ship trajectories have always less or more local character, and, that only spherical-symmetric form of the Earth has absolute character. Within standard quantum mechanical formalism collapse by measurement, i.e. detection must be postulated independently of the unitary symmetric (that conserves superposition or entanglement) quantum mechanical dynamics since to this day there is no realized experiment on the existence of the entanglement between quantum system and detector (Schrödinger cat effect). In this work we suggest an original, realistic and simple, form of the experiment of interference of single photon at beam splitter. Here quantum entanglement between photon and photon trajectories detector (which corresponds to single simple two-level system, practically single atom) is studied. (All this is equivalent to elementary processes within typical gas laser tube and in this sense suggested entanglement can be considered even experimentally verified.) It admits model of the collapse by detection as the spontaneous breaking of the unitary symmetry of quantum mechanical dynamics as an especial case of the general formalism of the spontaneous symmetry breaking. Collapse represents an effective, local phenomena, while unitary symmetry represents an absolute exact, global phenomena. It can be extremely important for the quantum mechanics foundation.

In early days of navigation over sea and ocean it has been considered that when a ship, governed by waves, arrives as some cri9tical distance of the global character, then this ship must absolute collapse, i.e. must fall in the nothing. Only remarkable Magellan voyage definitely "experimentally" realized proof that there is no absolute collapse of the ship trajectory on the ocean, that different collapses of the ship trajectories have always less or more local character, and, that only spherical-symmetric form of the Earth has absolute character. Within standard quantum mechanical formalism collapse by measurement, i.e. detection must be postulated independently of the unitary symmetric (that conserves superposition or entanglement) quantum mechanical dynamics since to this day there is no realized experiment on the existence of the entanglement between quantum system and detector (Schrödinger cat effect). In this work we shall suggest an original, realistic and simple, form of the experiment of interference of single photon at beam splitter. Here quantum entanglement between photon and photon trajectories detector (which corresponds to single simple two-level system, practically single atom) will be studied. (All this is equivalent to elementary processes within typical gas laser tube and in this sense suggested entanglement can be considered even experimentally verified.) It admits model of the collapse by detection as the spontaneous breaking of the unitary symmetry of quantum mechanical dynamics as an especial case of the general formalism of the spontaneous symmetry breaking. Collapse represents an effective, local phenomena, while unitary symmetry represents an absolute exact, global phenomena. It can be extremely important for the quantum mechanics foundation.

As it is well-known [1]-[6] experiment of the interference of single quantum system (photon, electron, etc.) at a diaphragm with two slits or, analogously, at beam splitter represents corner stone for demonstration of the basic principles of the standard quantum mechanical formalism [2], [3], [7]-[9]. In this experiment, simply speaking, source S emits single quantum system (with well-defined momentum or corresponding de Broglie wave-length) that propagates toward beam splitter BS. In time moment chosen to be initial quantum system quantum dynamically interacts with beam splitter. As the result of this interaction initial quantum state of the quantum system evolves deterministically during time in the following quantum superposition of quantum reflected state, i.e. reflected trajectory $|R(t)\rangle$ and quantum transmitted state, i.e. transmitted trajectory $|T(t)\rangle$

(1) $$i\, 2^{-1/2} |R(t)\rangle + 2^{-1/2} |T(t)\rangle$$

Reflected and transmitted state will be later redirected by two total mirrors MR and MT (at the same distance in respect to beam splitter) respectively toward the same detection photo plate DPP. For this reason it will be considered that total reflected trajectory is determined, roughly speaking, by centers of BS,MR and DPP, while total transmitted trajectory is determined by centers of BS,MT and DPP. In any individual case detection photo plate detects finally always only one photon. But at a statistical ensemble of the individual detections statistical distribution of the photons along detection photo plate axis can be different.

If between beam splitter and detection photo plate quantum system does not interact with any other physical system able to detect concrete trajectory of quantum system then detection photo plate finally detects statistical distribution with characteristic interference terms (existence of many local minimums and maximums of the distribution function) corresponding to superposition (1) (Figure 1.). It is very important to be pointed out that here detection photo plate realizes detection at the quantum system as the separated system.

But if between beam splitter and photo plate quantum system interacts with such physical system, called detector D, able to detect concrete trajectory of quantum system then photo plate finally effectively detects statistical distribution with effective absence of the interference terms (distribution function with one maximum and without any minimum): It corresponding to principally probabilistic breaking (collapse) of superposition (1) in statistical mixture of reflected and transmitted trajectory with statistical weights $|i\, 2^{-1/2}|^2$ and $|2^{-1/2}|^2$ respectively (Figure 2.). Such collapse cannot be exactly presented as the exact result of the deterministic quantum dynamical evolution. But by quantum dynamical interaction between quantum system and detector corresponding entanglement should be realized. Entanglement does not admit unambiguous separation of the quantum super-system (in entangled state) in its sub-systems (described by pure or statistically mixed quantum states). For this reason here detection photo plate realizes only effectively exact but not absolutely exact sub-systemic detections of the quantum system at entangled quantum super-system that includes both quantum system and detector. Even if existence of the entanglement is definitely experimentally proved in many cases, e.g. [10]-[12], existence of the entanglement between quantum system and detector (Schrödinger cat effect) has been never experimentally detected to this day. Now we shall suggest an original, realistic and simple, form of the experiment of interference of single photon at beam splitter. Here quantum entanglement (Schrödinger cat effect) between photon and photon trajectories detector (which corresponds to single simple two-level system, practically single atom) can be studied. (All this is equivalent to elementary processes within typical gas laser tube and in this sense suggested entanglement can be considered even experimentally verified.) It can be extremely important for the quantum mechanics foundation.

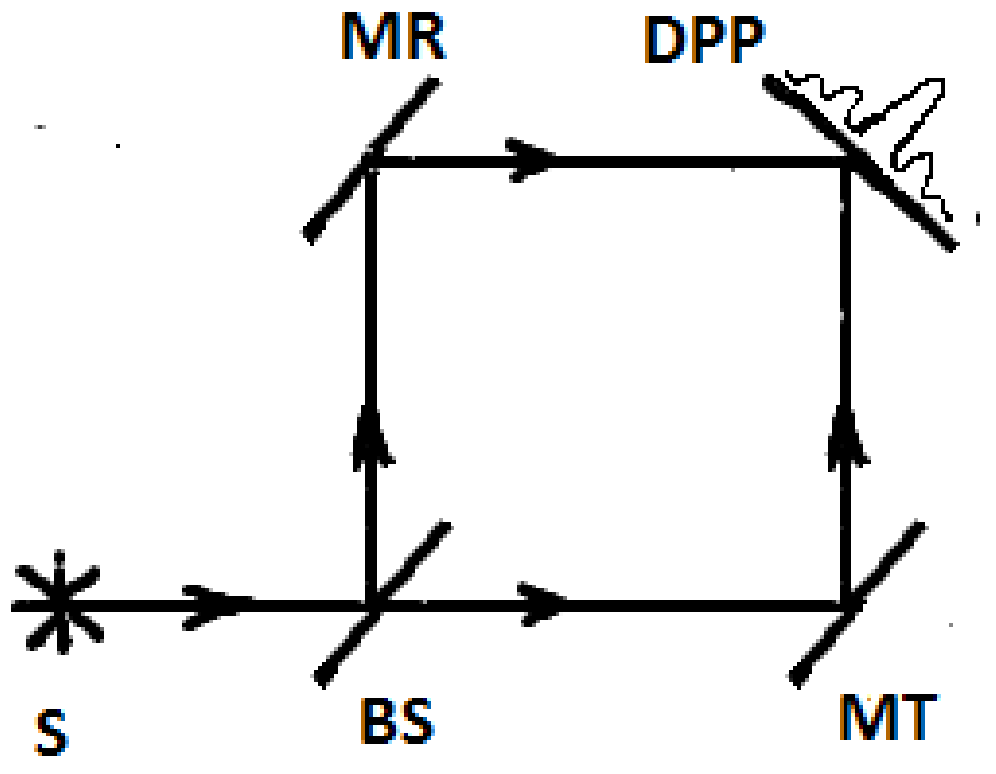

**Figure 1.**

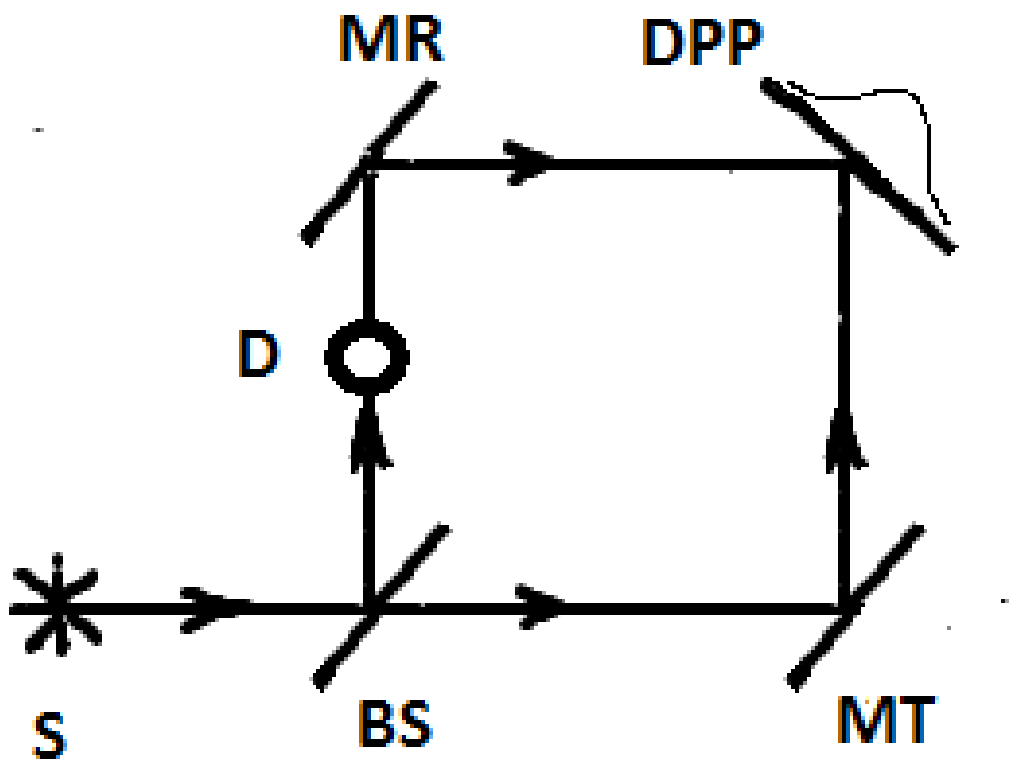

**Figure 2.**

As it has been pointed out by Feynman [3], [4] it is not necessary that detector be a macroscopic system. Namely, it is well-known that, for example, by interference of single electron at diaphragm with two slits, detector can be in principle even single photon, but of course technically is much simpler that experiment be realized with many photons, i.e. macroscopic ensemble of the photons.

Suppose that quantum system corresponds to single photon P while photon trajectories detector corresponds to single simple two-level system TLS, practically single atom, with excited state |E> and ground state |G> (Figure 3.). Suppose that difference between energy of the excited state and energy of the ground state is equal to photon energy. Also suppose that life time of the excited state of two-level system is much larger than time propagation of the photon between source, beam splitter and detection photo plate. Finally suppose that motion of two-level system during photons propagation can be neglected effectively. Such TLS in excited state will be initially placed, roughly speaking, at photon reflected trajectory in the middle between beam splitter and MR

If photon propagates along reflected trajectory its quantum dynamical interaction with two-level system TLS in excited state can be considered as the stimulated emission of a new, copy photon coherent with initial, without any change of the reflected state of the initial photon. Simultaneously TLS turns out in the ground state. Later two photons that propagate coherently along reflected trajectory, after redirection by MR will be finally detected by detection photo plate.

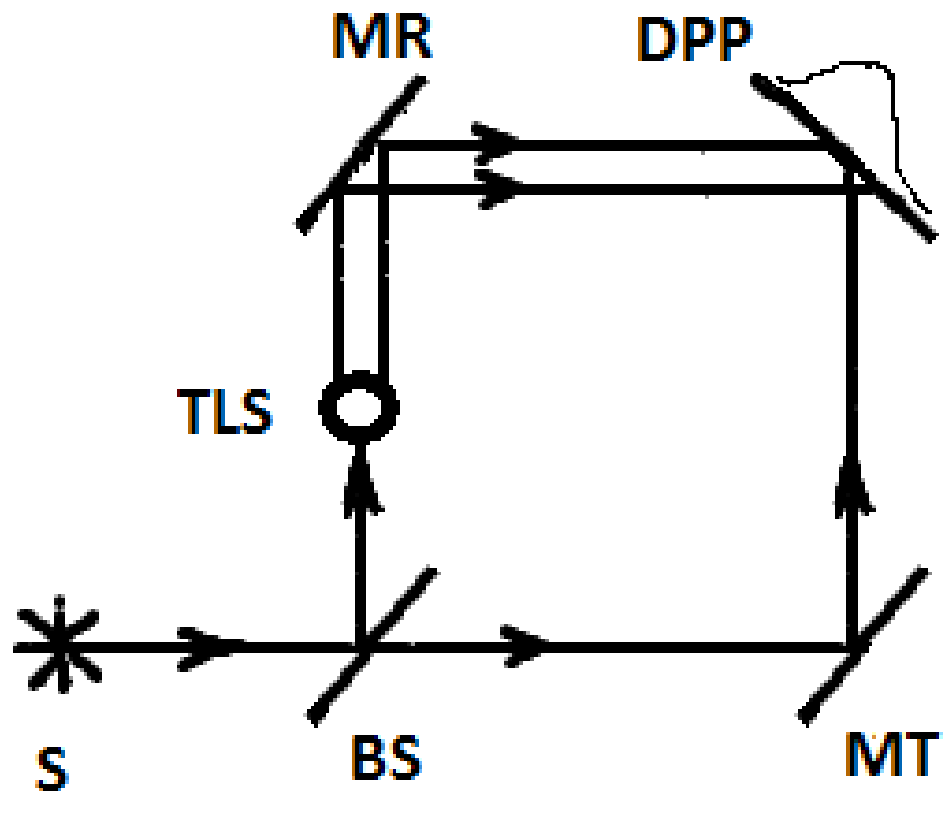

Figure 3.

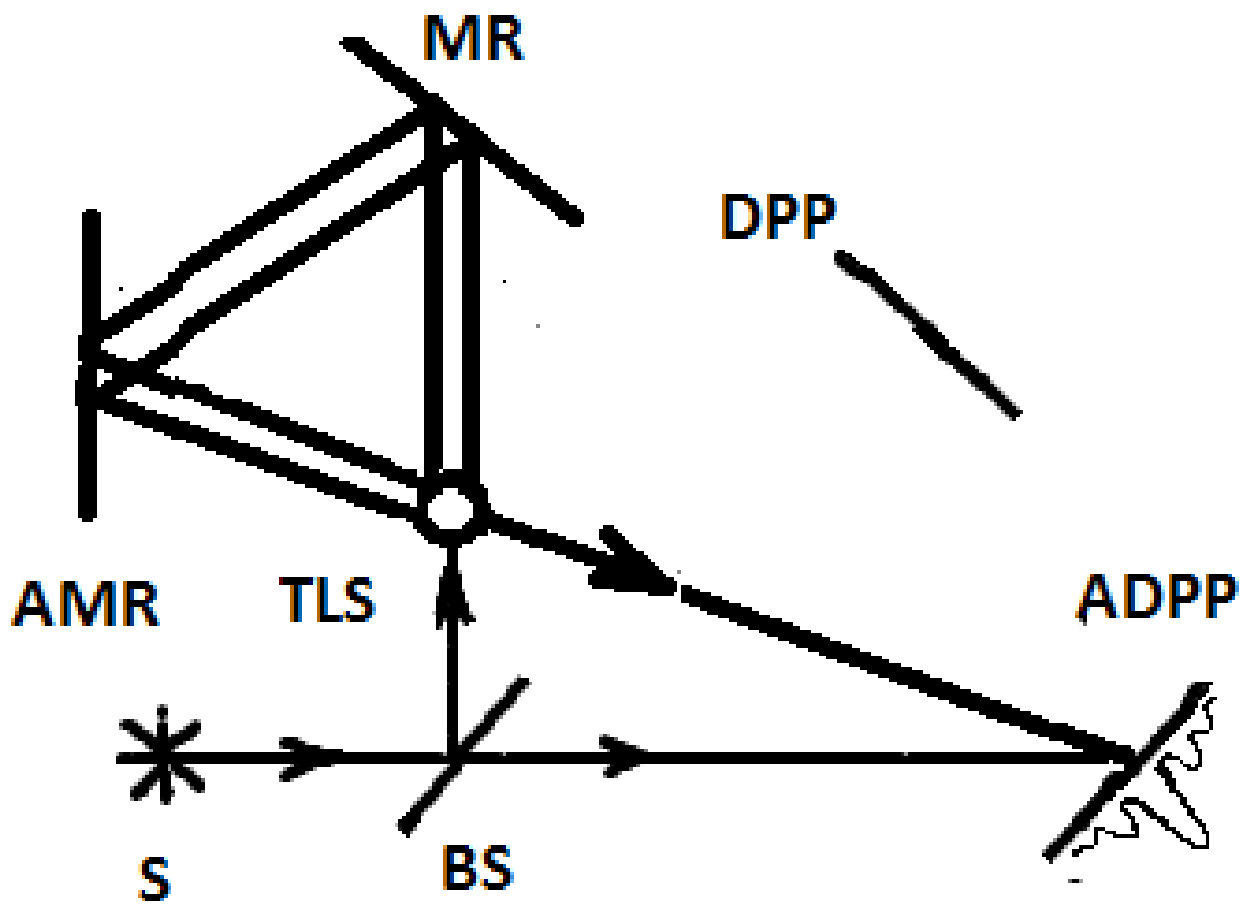

Figure 4.

If photon propagates along transmitted trajectory it quantum dynamical does not interact with two-level system that stands in the initial excited state. Finally, after redirection by total mirror MT, photon propagates toward detection photo plate DPP where it will be finally detected.

In this way it is definitely proved that mentioned two-level system can be consistently treated as detector of the photon trajectories. Also it can be repeated and pointed out the following. By mentioned detection neither reflected nor transmitted trajectory of the photon will be changed.

But, as it has been previously discussed, before quantum dynamical interaction between P and TLS, for P in superposition state (1) and TLS in excited state, quantum super-system P+TLS (that includes both, P and TLS) is described by the following non-

entangled quantum state

(2) $( i\, 2^{-1/2} |R(t)> + 2^{-1/2} |T(t)>) |E>$

Immediately after quantum dynamical interaction between P and TLS non-entangled quantum state (2) deterministically evolves in the following, entangled state

(3) $i\, 2^{-1/2} |R(t)> |R(t)> |G> + 2^{-1/2} |T(t)> |E>$
.

First term of (3) describes two identical photons that propagate along the same reflected trajectory toward MR and TLS in ground state, while second term of (3) describes photon that propagates along transmitted trajectory toward MT and TLS in excited state.

Immediately after quantum dynamical interactions between photons and MR and MT entangled quantum state (3) deterministically evolves in the following, entangled state

(4) $i\, 2^{-1/2} (i |R(t)>)(i |R(t)>) |G> + 2^{-1/2} (i |T(t)>) |E>$
.

First term of (3) describes two identical photons that propagate along the same reflected trajectory toward DPP and TLS in ground state, while second term of (3) describes photon that propagates along transmitted trajectory toward DPP and TLS in excited state.

Finally, detection photo plate, that realizes sub-systemic detection at the photons, will effectively exactly detects statistical mixture of two photons in reflected state and one photon in transmitted trajectory with corresponding statistical weights $|i\, 2^{-1/2}|^2$ and $|2^{-1/2}|^2$ respectively. Moreover, by additional experimental analysis it can be shown that in any individual case of the detection two photons detected by DPP corresponds to TLS in ground state, while one photon detected by DPP corresponds to TLS in excited state. Simply speaking DPP certainly repeats results of the TLS detections.

Existence of the entangled state (3) can be simply experimentally te4stified by experimental arrangement presented at Figure 4. . Here, instead of MT, an additional detection photo plate ADPP, is placed. Also, an additional total mirror for reflected photon AMR is introduced. Finally, MR is rotated in such way so that it redirects reflected photon toward AMR and so that AMR redirects the same photon toward TLS and ADPP. In this way a new, additional reflected quantum state or additional reflected trajectory |AR(t)> is introduced and it is, roughly speaking, defined by centers of MR, AMR, TLS and ADPP. As it is not hard to see, all mentioned changes of the experimental arrangement does not change entangled state (3) that is realized by the first quantum dynamical interaction between P and TLS.

Immediately after quantum dynamical interactions between photons and MR entangled quantum state (3) deterministically evolves in the following, entangled state

(5) $i\, 2^{-1/2} (i |AR(t)>)(i |AR(t)>) |G> + 2^{-1/2} |T(t)> |E>$

First term of (5) describes two identical photons that propagate along the same additional reflected trajectory toward AMR and TLS in ground state, while second term of (5) describes photon that propagates along transmitted trajectory toward ADPP and TLS in excited state.

Immediately after quantum dynamical interactions between photons and AMR entangled quantum state (5) deterministically evolves in the following, entangled state

$$(6) \qquad i\, 2^{-1/2} (i^2 |AR(t)>)(i^2 |AR(t)>) |G> + 2^{-1/2} |T(t)> |E> = i\, 2^{-1/2} |AR(t)> |AR(t)> |G> + 2^{-1/2} |T(t)> |E>$$

First term of (6) describes two identical photons that propagate along the same additional reflected trajectory toward TLS and TLS in ground state, while second term of (6) describes photon that propagates along transmitted trajectory toward ADPP and TLS in excited state.

Quantum dynamical interaction between two identical photons in the additional reflected state and TLS in ground state can be simply considered as the absorption of one photon by TLS that turns out in excited state. Simultaneously, other photon propagates in the same additional reflected state toward ADPP. For this reason, immediately after quantum dynamical interactions between photons and TLS entangled quantum state (6) deterministically evolves in the following, non-entangled state

$$(7) \qquad ( i\, 2^{-1/2} |AR(t)> + 2^{-1/2} |T(t)>) |E> .$$

Within this non-entangled state (7) TLS is excited state while photon propagates toward ADPP in superposition state.

$$(8) \qquad i\, 2^{-1/2} |AR(t)> + 2^{-1/2} |T(t)>$$

This state (8) (i.e. corresponding interference terms) can be simply detected by final detection by ADPP. Of course, existence of (8) unambiguously implies existence of the entangled state (3). Until realization of mentioned experiment of the existence of the photon-two-level systems entanglement it can be expected that existence of mentioned entanglement represents the best chance for the Nature. (In fact we paraphrase Einstein words on the expectations of the results of the first experiments, i.e. astronomical observations that testify general theory of relativity.) Moreover, since as it is not hard to see, quantum dynamical interactions between photon and two level system are equivalent to elementary processes within typical gas laser tube, suggested photon-detector entanglement can be in this sense considered even experimentally verified.

Thus, if entanglement between photon and TLS as photon trajectories detector really exists, then additional formalization of the collapse by detection is necessary. For this reason we shall apply model of the (self-)collapse at detector as the spontaneous (non-dynamical) breaking (effective hiding) of the unitary symmetry (superposition or entanglement) of quantum mechanical dynamics [13],[14]. It can be considered as an especial case of the general formalism of the spontaneous symmetry breaking [15]-[17].

Consider again entangled state (3). First term of (3) $|R(t)> |R(t)> |G>$ represents product of three quantum states while second term of (3) $|T(t)> |E>$ represents, formally, product of two quantum states. It needs an additional argumentation. Firstly it is well-known that for time independent Hamiltonian observable of quantum system its eigen values and eigen states are time independent. It implies, according to standard quantum mechanical dynamics, that such quantum system (without interactions with other physical systems), initially prepared in some excited quantum state, must stand in the same excited quantum state in any later time moment. But as it is well-known in the real experimental situations quantum system, by means of spontaneous emission of single photon, turns out from excited in some less excited or ground state after relatively short time interval, i.e. life time. Thus within standard quantum mechanical dynamics for two-level system it is supposed that excited state |E> and ground state |G> represent two state from the same Hilbert space. After spontaneous photon emission excited state |E> turns out in state |P>|G> where |P> represents state of the spontaneously emitted photon. But obviously |P>|G> and |G> do not represent states in the same Hilbert space, which means that exactly speaking |E> and |G> do not represent states in the same Hilbert space. Moreover, it implies that excited state before spontaneous photon emission represents state of interacting two-level system and electromagnetic field formally denoted by |TLS,P (t)> whose exact dynamical evolution needs extension of the standard quantum mechanical dynamics by quantum field theory dynamics. But such extension can be approximately neglected in most usual quantum mechanical considerations [7]. But, in some examples, as it is mentioned form of the entangled state (3) such approximation is not sufficient. For this reason we must here use somewhat better approximation during initial stability of the excited state. Within such approximation |E> can be changed by |P,TLS> |G> where |P,TLS> represents formally quantum state of the electromagnetic field, i.e. photon localized within domain of TLS. Now we can rewrite (3) in the following, more correct form

(9) $$i\, 2^{-1/2} |R(t)> |R(t)> |G> + 2^{-1/2} |T(t)> |P,TLS1>|G> = ( i\, 2^{-1/2} |R(t)> |R(t)> + 2^{-1/2} |T(t)> |P,TLS1>)|G> \quad .$$

State (9) is partially entangled. Namely here TLS in ground state is formally separated from two entangled photons in entangled state

(10) $$i\, 2^{-1/2} |R(t)> |R(t)> + 2^{-1/2} |T(t)> |P,TLS1> \quad .$$

Entangled state (10), simply speaking, describes entanglement of two photons so that first entanglement term describes both photons that free propagate along reflected trajectory while second entanglement term describes one photon that free propagates along transmitted trajectory and other photon localized within TLS. More accurately, entangled state (10) correlates two eigen states, reflected and transmitted, of the first, free propagating photon, and, two eigen states of the second photon first of which is reflected characteristric for free moving photon and second of which is characteristic for photon localized within TLS. Reflected and transmitted state of the first photon weakly interfere by photon propagation except in small vicinity of the detection photo plate where mentioned states become strongly interfering. On the other hand reflected state and localized (within TLS) state of the other photon are practically always extremely weakly interfering.

Consider quantum mechanics as exact physical theory and its unitary symmetric dynamics (that conserves superposition or entanglement) as the exact dynamics.

Suppose that quantum dynamical state represents a wave packet. Under additional, well-known [2],[7], [18] approximation conditions mentioned wave packet can be globally approximately treated as the classical mechanical particle that satisfy approximate classical mechanical dynamical equation. (Namely, in Ehrenfest picture, average value of the quantum dynamical equation can be Taylor expanded [18] so that first term in expansion represents classical dynamical term for average coordinate value while other terms are proportional to increasing degrees of the coordinate uncertainty exponents. If coordinate average value is much larger than coordinate uncertainty, i.e. wave packet width, first term turns out in classical dynamical term for wave packet while other terms can be effectively neglected. Down limit of such wave packet approximation is characterized by Heisenberg uncertainty relations.) In this way (until all approximation conditions are satisfied) classical mechanics can be considered as approximate physical theory.

Consider exactly quantum mechanically two wave packets with practically the same (coordinate intervals) widths. It can be considered that mentioned two wave packets weakly interfere if distance between their centers is larger than one width.

Consider quantum dynamical state that represents (non-trivial) superposition of two weakly interfering wave packets. As it has been proved in [13],[14], such quantum superposition cannot be globally classical presented as the classical particle. (Namely, then mentioned Taylor expansion of Ehrenfest average value of the quantum dynamical equation becomes divergent.) Simply speaking, exactly existing superposition of two weakly interfering wave packets, approximately classically is globally non-stable and non-observable. But, of course, mentioned superposition is approximately classically locally stable and observable within any of two wave packets. Then, according to general formalism of the spontaneous symmetry breaking, here inherently probabilistically (with typical quantum mechanical probabilities) and spontaneously event of appearance of one or other wave packet becomes realized at classical level of analysis accuracy. However, at the quantum level of analysis accuracy superposition of weakly interfering wave packets stands conserved. For this reason mentioned event appearance does not correspond to any real “choice” if this word “choice” should imply deterministic description of the event within dynamical superposition breaking. However, if by new detection only such quantum observables for which mentioned wave packets represents eigen states will be analyzed, classical level can be used as technically simple theoretical method, but it cannot be considered as any principal “choice”. Then, according to determinism of the classical dynamics, new detector will detect the same wave packet which appears by the previous spontaneous superposition breaking. But if by new detection other (complementary) observables will be detected quantum superposition of weakly interfering wave packets must be used. Obviously all this represents an excellent model of the self-collapse at classical level of analysis accuracy.

Further, consider quantum dynamical state that represents (non-trivial) superposition of two wave packets. Suppose that initially mentioned wave packets are strongly interfering but that, according to deterministic quantum dynamical evolution, distance between centers of wave packets increases during time. Then in moment when wave packets centers become sufficiently distant, i.e. when wave packets become weakly interfering, conditions for self-collapse as spontaneous superposition breaking become satisfied and in this sense we can speak about collapse as a continuous phase transition.

Consider now measurement or detection procedure. Before quantum dynamical interaction between quantum system and detector, quantum system is described by superposition of the eigen states of measured observable, while detector is described by ”zero” eigen state of the pointer observable. During deterministic quantum dynamical interaction between quantum system and detector entangled quantum state (bi-orthogonally expanded over quantum system measured observable eigen states and detector pointer observable eigen states) of the quantum super-system (that includes quantum system and detector) becomes realized. This entangled state (in absence of new interactions with additional physical systems) stands conserved during time. Suppose that within mentioned entangled state of super-system detector pointer observable eigen states represent wave packets. Suppose that mentioned wave packets are initially strongly interfering and that mentioned wave packets become during time weakly interfering. In moment when all mentioned wave packets become mutually weakly interfering at the classically described detector self-collapse as spontaneous (non-dynamical) entanglement breaking (effective hiding) appears. In other word, here spontaneous superposition breaking at super-system appears in full analogy with spontaneous superposition breaking at simple system. Simultaneously, for reason of correlations between detector pointer observable eigen states and quantum system measured observable eigen states, in respect to self-collapsed detector quantum system becomes effectively uniquely described by corresponding statistical mixture of the eigen states of measured observable. It this sense at quantum system relative collapse effectively appears as seemingly “absolute”. But within exact entangled state of super-system statistical mixture by relative collapse of the quantum system is one of many possible second kind mixtures of this quantum system. Relative collapse at quantum system is effective but not absolute quantum phenomena and it occurs only in respect to classically self-collapsed detector. Super-system that includes quantum system and detector is exactly described by entangled state. If by a new sub-systemic detection at quantum system only, only such quantum observables compatible with previously detected observable will be analyzed, quantum system can be effectively exactly described by statistical mixture characteristic for previous relative collapse. In this sense detections of the first and new detector are identical. But if by new detection other (complementary) observables will be detected entangled state of super-system, i.e. a different second kind mixture of the quantum system must be used. Obviously all this represents an excellent model of the measurement or detection if full agreement with all known experimental data.

Previously discussed model we shall now apply for explanation of previously suggested experiment of detection presented at Figure 3.

In mentioned experiments excited state of two-level system is approximated with product of the state of single photon localized within mentioned two-level system and ground state of two-level system. We can additionally consistently approximate state of the photon localized within a two-level system by a wave packet with coordinate corresponding to two-level system position (and width corresponding to linear dimension of the two-level system) and with zero momentum (since it has been supposed that two-level system motion can be approximately neglected).

Now we can again consider exact entangled state of two photons (10) realized immediately after quantum dynamical interactions between photon and two-level system. In this entangled state quantum system represents first, free propagating photon with two eigen states or two eigen trajectories, reflected and transmitted. Simultaneously, in this entangled state detector represents other photon with first eigen state corresponding to photon localized within TLS and second eigen state corresponding to free propagating transmitted photon. First and second detector eigen states always extremely weakly interfere, but here only first eigen state represents wave packet. Nevertheless, absorption of the this second eigen state, i.e. free propagating photon by some micro-element, e.g. some atom of the detection photo plate, this second eigen state becomes also localized. Then it can be additionally approximated by a wave packet with coordinate corresponding to position of mentioned micro-element of detection photo plate and with zero momentum. In this way all condition for the realization of the self-collapse at mentioned detector, i.e. localized photon become satisfied and this self-collapse occurs as the spontaneous (non-dynamical) unitary symmetry (superposition or entanglement) breaking (effective hiding) at the approximate classical level of analysis accuracy. Simultaneously, in respect to self-collapsed localized photon, at free propagating photon relative collapse as effective but not absolute exact quantum phenomena occurs. Absolutely exact is only entangled state (9) at the exact quantum level of analysis accuracy.

In conclusion the following can be repeated and pointed out. In early days of navigation over sea and ocean it has been considered that when a ship, governed by waves, arrives as some critical distance of the global character, then this ship must absolute collapse, i.e. must fall in the nothing. Only remarkable Magellan voyage definitely "experimentally" realized proof that there is no absolute collapse of the ship trajectory on the ocean, that different collapses of the ship trajectories have always less or more local character, and, that only spherical-symmetric form of the Earth has absolute character. Within standard quantum mechanical formalism collapse by measurement, i.e. detection must be postulated independently of the unitary symmetric (that conserves superposition or entanglement) quantum mechanical dynamics since to this day there is no realized experiment on the existence of the entanglement between quantum system and detector (Schrödinger cat effect). In this work we suggest an original, realistic and simple, form of the experiment of interference of single photon at beam splitter. Here quantum entanglement between photon and photon trajectories detector (which corresponds to single simple two-level system, practically single atom) is studied. (All this is equivalent to elementary processes within typical gas laser tube and in this sense suggested entanglement can be considered even experimentally verified.) It admits model of the collapse by detection as the spontaneous breaking of the unitary symmetry of quantum mechanical dynamics as an especial case of the general formalism of the spontaneous symmetry breaking. Collapse represents an effective, local phenomena, while unitary symmetry represents an absolute exact, global phenomena. It can be extremely important for the quantum mechanics foundation.